\documentclass[final,3p,times]{elsarticle}

\usepackage{graphicx}

\usepackage{amssymb}

\journal{}

\begin{document}

\begin{frontmatter}

\title{Emergent quantum mechanics of finances}
\author{Vadim A.~Nastasiuk\corref{cor1}}
\ead{nasa@i.ua}
\cortext[cor1]{Tel.: +380487347896}
\address{South Ukrainian National Pedagogical University, Staroportofrankivska Str., 26, Odessa, Ukraine 65020}
\begin{abstract}
This paper is an attempt at understanding the quantum-like dynamics of financial markets in terms of non-differentiable price-time continuum having fractal properties. The main steps of this development are the statistical scaling, the non-differentiability hypothesis, and the equations of motion entailed by this hypothesis. From perspective of the proposed theory the dynamics of S$\&$P500 index are analyzed.
\end{abstract}

\begin{keyword}
econophysics \sep financial markets \sep statistical scaling \sep fractals \sep Schr\"{o}dinger equation


\end{keyword}

\end{frontmatter}


\section{Introduction}
In the last decades there were performed numerous investigations on applying quantum mechanics to financial markets, in particular secondary financial markets (see e.g. [1-8] and book \cite{BA}, and many publications in arxiv.org). Unfortunately, the constructing of quantum theory of finances often reduces to simple postulation of the Schr\"{o}dinger equation and its subsequent solution under some entry conditions. Meanwhile, the fundamental property of markets, which makes the price paths similar to trajectories of microscopic particles, namely, fractality is not taken into account. We would refer to the Feynman's studies \cite{FH} and results of work \cite{AW}, which have shown that the typical trajectories of quantum particles are continuous but non-differentiable, and can be characterized by a fractal dimension $D=2$ at small length scales. It is well known that this is a fractal dimension of Brownian motion or, from the mathematical point of view, the Wiener process. In his time Nelson assumed \cite{NL} any particle to be subjected to an underlying Brownian motion of unknown origin. This hypothesis led him in a natural way to the Schr\"{o}dinger equation but with entirely classical physical interpretation. Later Nottale developing his idea of scale invariance of physical laws in series of works [13-16] and book \cite{N5}, attributed the property of fractality to very space-time continuum, and thus recover the quantum mechanics as a manifestation of space non-differentiability at small scales.

In financial mathematics it is widely accepted that price time series constitute a some deformation of the Wiener process. Keeping in mind the Nelson's hypothesis we can suppose correctly defined price trajectories relying to some mechanical system, and try to reveal the laws of dynamics for it. Retracing the main steps of Nottale's formalism we  show in this paper that price dynamics is of quantum character and the master equation is the Shr\"{o}dinger equation.

 \section{Statistical scaling}
The financial markets have such property as statistical scaling or fractality reflected in that their parts 'is the same' as the structure of the whole object. For instance, if the logarithmic returns (differences of logarithms of price) taken for equal time intervals $\tau$, have probability density $p(\xi,\tau)$, then the values of logarithmic returns taken for time intervals $k\tau$, $k > 0$, will be subject to the distribution \cite{SY}
\begin{equation} \label{1} p(k^H\xi,k\tau)=k^{-H}p(\xi,\tau).
\end{equation}

Here $H$ is the Hurst exponent, reciprocal value which, $D = 1/H$ is named statistical fractal dimension.

The scaling (\ref{1}) is observed for a wide time range on most of markets \cite{MS}.

When the number $H$ is estimated being close to $1/2$ $(D = 2)$ the \textit{geometric} (\textit{economic})  \textit{random walking approximation} \cite{SA} is valid: the sequence of logarithmic price returns is appeared to be a set of independent normally distributed random numbers with zero mean and standard deviation (volatility) $\propto\tau^H$. A continuous modification of that process is known as Brownian motion or Wiener process. This model corresponds to the famous \textit{efficient market hypothesis} which has its basis in the presumption that prices simultaneously absorb all current economical information and change only when new information comes out.

Nowadays the geometric Brownian motion is considered only as a first approximation of what is observed in real data. The exponent $H$ shows a systematic exceeding of $1/2$ ($D<2$) in many markets \cite{PE}, consequently the empirical volatility demonstrates a some more strong dependence on $\tau$, that usually associated with the emergence of a 'memory' or persistence in price series. The memory effects relevant rather to the absolute values (or squares) of price decrements which occur positively correlated. The presence of long-range correlations in the nonlinear functions of price changes indicates that there has to be one more fundamental (probably stochastic) process, in addition to the price change itself, which is usually attributed to volatility. The non-stationary of volatility explains to some extend the high leptokurtosis (narrow and large maximum and fat tails) of empirical distributions.

The overview of relevant statistical models is given in the book \cite{MS}. There is no general consensus on question, which stochastic process describes more adequately financial markets. But it is clear that in any case the scale should be present in the basic equation of price motion that one intends to construct.
\section{Non-differentiability}

The current position of price described by continuous function of time $x(t)$ is defined by a set of successive decrements
\begin{equation}\label{Wp}dX = dx + d\xi
\end{equation}
to be written in terms of its mean, $dx = <dX>$ and fluctuation respective to the mean $d\xi$ (such that $<d\xi> = 0$). An explicit dependence on scale (i.e. time resolution) is accounted for by description of the fluctuations. According to Section 2 and to Wiener's definition of the random function \cite{KS} that dependence is
\begin{equation}\label{3}
d\xi \propto dt^{1/D}.
\end{equation}
where the differential element $dt$ is identified with time resolution $\tau$.

This brings us back to the known problem of non-differentiability of Brownian trajectories \cite{SY,KS}. The two definitions of derivative
$$\frac{d}{dt} f(t)=\lim_{dt\to 0^+}\frac{f(t+dt)-f(t)}{dt}=\lim_{dt\to 0^+}\frac{f(t)-f(t-dt)}{dt},$$
which are equivalent in the differentiable case, in the non-differentiable situation fail since the limits are no longer defined. To discuss the kinematics of stochastic processes, therefore, we need a substitute for the derivative.

In the framework of scale relativity \cite{N3} the standard function $f(t)$ is replaced by a fractal function $f(t,dt)$ explicitly dependent on the time-resolution interval. As a consequence, one is led to define the two new derivatives
 $$
 \frac{d_+}{dt} f(t,dt)=\frac{f(t+dt)-f(t)}{dt}, \frac{d_-}{dt} f(t)=\frac{f(t)-f(t-dt)}{dt},
$$
where the transition $dt\rightarrow 0$ is still considered but without going to the limit $dt=0$ (which is now undefined) as in standard calculus. When applied to the price coordinate, these definitions yield a couple of fractal velocities  $d_{\pm}X(t,dt)/dt$, each decomposing in terms of mean part
\begin{equation}\label{d2}
 v_{\pm}=\left<\frac{d_{\pm}X}{dt}\right>,
\end{equation}
which are differentiable and independent of resolution, and scale-dependent fluctuation part.

There is no reason \emph{a priori} for the two classical velocities $v_{\pm}$ be equal. So, putting Eqs. (\ref{Wp}), (\ref{3}) and (\ref{d2}) together, we get into consideration two different stochastic processes
$$ dX_{\pm}=v_{\pm}dt + d\xi_{\pm}.
$$
In both the $d\xi_±$'s are supposed to have a mean of zero, mutually independent, and such that
\begin{equation} \label{d} \langle d\xi^2_{\pm}\rangle=\pm 2\mathcal{D} (dt)^{2/D},
\end{equation}
where $\mathcal{D}$ is a diffusion coefficient dimensionally. (Note, the particular type of statistical distribution (\ref{1}) is not essential; we suppose only the finiteness of its second moment).

The information needed to describe the system is therefore doubled with respect to the standard description. Proposed by Nottale \cite{N2} simple way to account for this doubling is to use complex numbers and combine the two derivatives (\ref{d2}) in a complex derivative operator,
\begin{equation}\label{nd}
\frac{\hat d}{dt}=\left<\frac{(d_+ +d_- )-i(d_+ - d_- )}{2dt}\right>,
\end{equation}
Applying operator of Eq. (\ref{nd}) to the position vector yields a complex velocity
$$\mathcal{V}=\frac{\hat d}{dt}x(t)=V-iU=\frac{v_+ +v_- }2 -i\frac{v_+ -v_- }2 .
$$

The real part $V$ of value $\mathcal{V}$ represents the classical mean velocity, while its imaginary part, $U$, is a new quantity arising from the non-differentiability.

In the special case of $D = 2$ we can take advantage of known It\^{o}'s lemma \cite{IT} and, expanding the differential of some smooth function $f(x(t), t)$ to second order, compute after averaging the pair of derivatives:
\begin{equation}\label{8}
\left<\frac{d_{\pm}f}{dt}\right>=\left(\frac{\partial}{\partial t}+v_\pm \nabla \pm \mathcal D \Delta\right)f.
\end{equation}
Here, for one-dimension price space the symbol $\nabla$  and $\Delta$ simply denote first and second derivative respectively.

On next step, substituting (\ref{8}) into (\ref{nd}), one gets the final expression for a time derivative:
\begin{equation}\label{vstav}
  \frac{\hat d}{dt}=\frac {\partial}{\partial t}+\mathcal V \nabla-i\mathcal D \Delta.
\end{equation}

As one can see an accounting of the non-differentiability entails the replacement of standard time derivative by new complex operator (\ref{vstav}). Note that the pure second derivative operator in the term $\mathcal D\Delta f$ is a consequence of the fractal dimension two. In the case of $D \neq 2$ which will be covered in Section 7, instead one gets an explicitely scale-dependent behavior: $\mathcal D\tau^{(2/D)-1}\Delta f$ \cite{N1}.
\section{Financial Schr\"{o}dinger equation}
Now introducing a force function $\Phi$ we write the equation of motion is the Newton-Euler form:
\begin{equation}\label{9}\frac{\hat d}{dt}\mathcal V \equiv \left(\frac{\partial}{\partial t}+\mathcal V \nabla-i\mathcal D \Delta\right) \mathcal V = -\frac 1{m}\nabla\Phi.
\end{equation}

By doing this we assume tacitly that the price dynamics is Newtonian, and the least-action principle though taken for the complex Lagrangian \cite{N1} is fair. We suppose also that the non-random function $\Phi$ exists and describes influence of the conservative forces. Those forces have their sources in the global economic environment and describe the conditions in which specified market operates.

At next step we could separate Eq. (\ref{9}) into real and imaginary parts:
$$m\left(\frac{\partial}{\partial t}V-\mathcal D\Delta U+(V\cdot\nabla)V-(U\cdot\nabla)U\right)=-\nabla\Phi,$$
$$m\left(\frac{\partial}{\partial t}U-\mathcal D\Delta V+(V\cdot\nabla)U+(U\cdot\nabla)V\right)=0$$
and solve in principle the obtained system of 'hydrodynamic' equations under given real quantity $\Phi$, but we know nothing about possible boundary conditions.

Following Nottale we make substitution
\begin{equation}\label{10} \mathcal V =-2i\mathcal D \nabla(\ln\psi)
\end{equation}
in (\ref{9}) and after some algebraic transformations \cite{N1} come to the equation
\begin{equation}\label{100}\mathcal D^2\Delta\psi+i\mathcal D\frac{\partial}{\partial t}\psi-\frac{\Phi}{2m}\psi=0
\end{equation}
which, if one replaces $\mathcal D$ by $\hbar/2m$, takes the standard Schr\"{o}dinger equation form:
\begin{equation}\label{sch} i\hbar\frac{\partial\psi}{\partial t}=-\frac{\hbar^2}{2m}\Delta\psi+\Phi\psi.
\end{equation}

The obtained expressions (\ref{100}) and (\ref{sch}) describe the price dynamics in terms of the potential $\ln\psi$ for complex velocity field. The possibility of introducing of the potential is itself a consequence of assumption of fundamental financial field $\Phi$ potentiality.

The relation
\begin{equation}\label{12} \mathcal D=\hbar/2m
\end{equation}
allows us to take into consideration the asset inertial mass $m$ which is inversely proportional to diffusion. The constant $\hbar$ here is simply dimensional aspect ratio.

It is very difficult to give an unambiguous interpretation in financial terms to $m$. We emphasize only that the low volatility (big 'mass') involves the high liquidity. The market liquidity is asset's ability to be sold without causing a significant movement in price and with minimum loss in value. The price jumps be minor - the market be stable, when the buy orders volume and sell orders volume are nearly balanced. To do this, participants with the different investment horizon must be present. Because the horizons are different and vary by a factor of one million, economic agents take different decisions (to buy or to sell). If the market sheds investors with various investment horizons it loses liquidity and stability. There are no numerical characteristics for market liquidity or stability in modern financial theory. A candidate for this role could be the mass.

According to (\ref{d}) the mean quadratic velocity of price fluctuations is $\langle v^2\rangle\approx \langle(\delta x/\delta t)^2\rangle=2\mathcal D/\delta t$ and the corresponding energy is written as $\varepsilon = m\langle v^2\rangle/2\approx m\mathcal D /\delta t$. Taking into account (\ref{12}) and associating time interval $\delta t$ with a time resolution $\tau$ we derive the financial uncertainly relation:
\begin{equation}\label{13}\varepsilon\cdot\tau\approx \hbar/2.
\end{equation}

So the fluctuation energy inversely proportional to time interval. The passage to zero time resolution interval is devoid of physical meaning in quantum mechanics (an infinite energy would be needed to perform a measurement). When observing the markets we are faced with a somewhat different problem. With the continuous ($\tau\rightarrow 0$)  observation the price looks completely different. Solid trajectory disappears and we can see only ticks - (almost) instant (with $v\rightarrow\infty$) changing. Another words, the price keeps its value for some time, then it drops. This high- frequency pattern of the behavior of financial indexes disappears and we come back to the continuous paths under transition to finite periods $\tau >0$.

The ticks statistics requires another consideration \cite{SY} and don't analyzed here.
\section{Back to classical mechanics}
In present approach, we have obtained the Schr\"{o}dinger equation without introducing a probability density. Now our theory must be completed by the statistical interpretation.

The inverse substitution of the $\psi$-function of the form
\begin{equation}\label{psi} \psi=A\,\exp\left(\frac i{\hbar}S\right)
\end{equation}
into Eq.(\ref{10}) gives
\begin{equation}\label{14}
    V=\nabla S/m
\end{equation}
that makes possible interpreting $S$ as classical action. The same replacement of variables in the Schr\"{o}dinger Eq. (\ref{sch}) generates the Bohm-Madelung system of equations \cite{BH, HO}:
$$\frac{\partial A}{\partial t}=-\frac 1{2m}\left[A\nabla S+2\nabla A\cdot\nabla S\right],
$$
\begin{equation}\label{15}
    -\frac{\partial S}{\partial t}=\frac{(\nabla S)^2}{2m}+\Phi-\frac{\hbar^2}{2m}\frac{\Delta A}A.
\end{equation}

The first of Eqs. (\ref{15}) due to (\ref{14}) is transformed into continuity equation:
\begin{equation}\label{16}\frac{\partial P}{\partial t}=-\nabla(P\cdot V)
\end{equation}
for the function $P(x,t)=A^2=\psi\psi^*$ which may be interpreted as probability density of finding price spot in a certain place of one-dimensional price space. According to (\ref{16}) the probability density moves with a current velocity $V=\nabla S/m$.

The second ratio of (\ref{15}) is nothing but Hamilton-Jacobi equation with embedded quantum potential in the Bohmian form \cite{BH}:
$$Q=-\frac{\hbar^2}{2m}\frac{\Delta A}A.$$

In contrast to external force function $\Phi$ the quantum potential depending on probability amplitude has to reflect the influence of traders' expectations  or 'mental $\psi$-field' in interpretation of \cite{XR} and \cite{C}. Due to modern electronic communications the financial agents being almost instantly informed of all price movements may change their mind about a market, i.e. perform transactions that in turn affects on quotes. The corpuscular-wave duality in Bohmian interpretation is a quantum analogue of such recursive interaction.

We can give also quite a 'classical' interpretation to potential $Q$. Being expressed through the function
$$U=\mathcal D\nabla P/P
$$
it takes the form
$$Q=-\frac{mU^2}2-m\mathcal D\nabla U
$$
with average value
$$\langle Q\rangle=\int{\frac1 2 mU^2Pdx}$$
that has the form of energy, additional to mean external potential. According to Einstein's theory of Brownian motion (see \cite{NL}) this type of energy acquired by a particle, in equilibrium with respect to an external force, to balance the osmotic force. So, we can call $U$ the osmotic velocity. In our financial interpretation the osmotic pressure is created by imbalance in whole mass of traders' orders.
\section{Financial potential}
The global financial potential $\Phi$ may apparently be considered constant for long periods of time. If it is so, we have the stationary quantum mechanical task with $-\partial S/\partial t=E=const$ and the mean velocity $V =\nabla S/m$ being equal to zero. Then Schr\"{o}dinger equation reduces to
\begin{equation}\label{18}\frac 1 m (\Phi-E)=2\mathcal D ^2\frac{\Delta\sqrt P}{\sqrt P}.
\end{equation}

At this point we get opportunity to extract the potential $\Phi(x)$ (or the same, estimate the osmotic impact) from 'experimental' data. We take into consideration the S$\&$P500 index and calculate the probability $P(x)$ for the two time intervals: from 1997 to 2002 and from 2003 to 2008 (Fig. 1). \begin{figure}[hbt!]
\includegraphics[width=16cm,keepaspectratio=true]{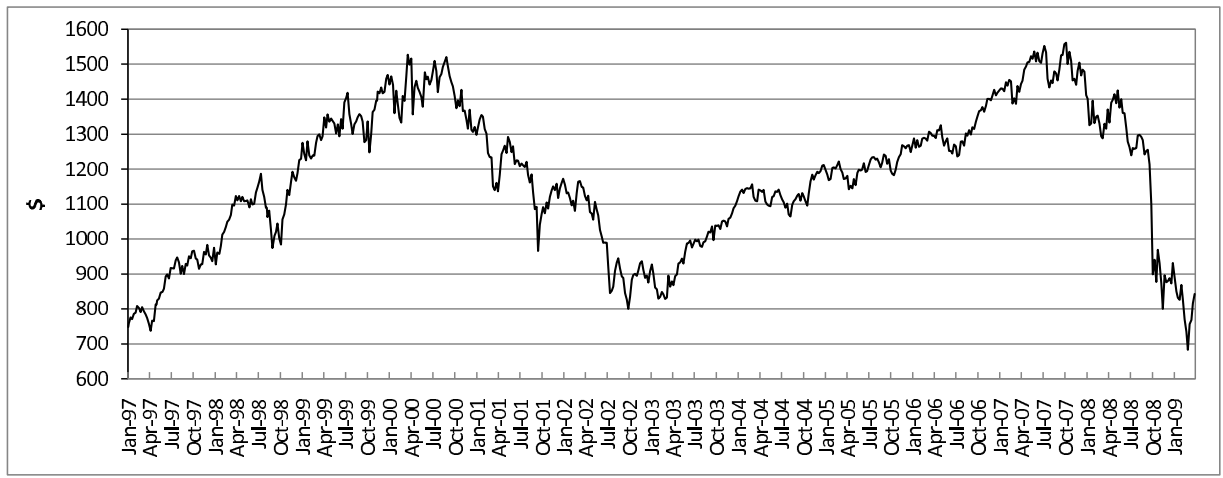}\\
\caption{S$\&$P500 index}\label{11}
\end{figure}
The selected periods cover long-term rises and declines and contain 327 and 311 trading weeks respectively. (The full data set is available at www.yahoo.com). The diffusion $\mathcal D$ during the specified periods is found to be equal 0,0169 and 0,0136 respectively.

To construct $P(x)$ we devide selected intervals into 19 pieces and count the number of timesteps the spot is in each specific box. On Figs. 2 and 3 the $x$ position is drawn horizontally, and the number of occurrences vertically. The histograms give the results of our counting.
 \begin{figure}[hbt!]
    \includegraphics[width=10cm]{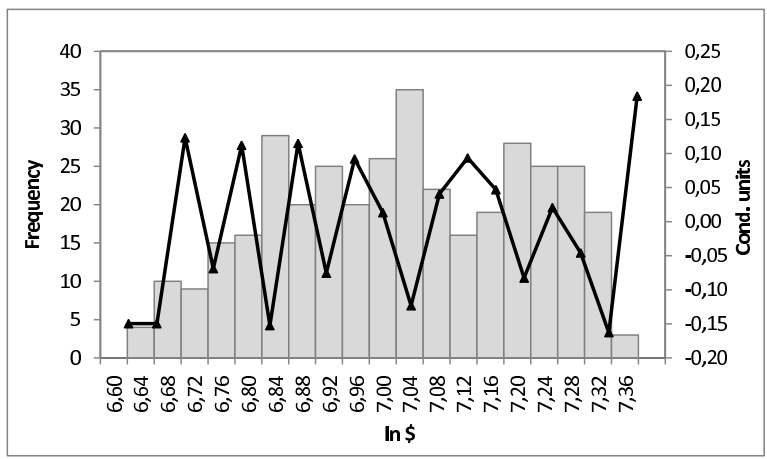}\\
  \caption{The function $2\mathcal D ^2\Delta\sqrt P/\sqrt P$ (solid line) for S$\&$P 500 over the period 1997-2002 versus corresponding distribution $P(x)$ (histogram). The second derivative of probability squire root is calculated as $y"=(x_{k+1}-2x_k+x_{k-1})/\Delta x^2$, with $\Delta x=0.04$}\label{22}
\end{figure}
\begin{figure}[hbt!]
  \includegraphics[width=10cm]{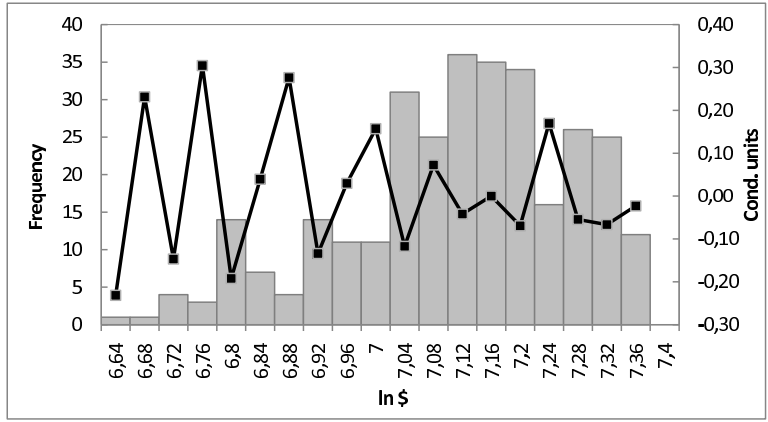}\\
  \caption{The function $2\mathcal D ^2\Delta\sqrt P/\sqrt P$ (solid line) for S$\&$P 500 over the period 2003-2008 versus corresponding distribution $P(x)$ (histogram). The second derivative of probability squire root is calculated as $y"=(x_{k+1}-2x_k+x_{k-1})/\Delta x^2$, with $\Delta x=0.04$}\label{33}
\end{figure}

As it is seen on Fig. 1 the spot does not go beyond the bounds of the price range (600; 1600)\$ during both periods of time. So we can assume the potential well. Then the result of substitution of $P$ into Eq. (\ref{18}) (solid lines on Figs. 2 and 3) should be considered as a rough estimate of the well form.

Here we solve in fact the return problem of quantum mechanics but cannot define the 'scattering' potential with whole accuracy because the number of tests is limited to one.

It is seen on Figs. 2 and 3 the oscillations of well bottom with period of 0,08 for both considered time intervals. The number of oscillation is approximately the same but the location of oscillation shifted. An identification of the same regularity in the potential function updates for future time periods might help in solving the problem of price prediction.
\section{Varying volatility and generalized Schr\"{o}dinger equation}
Now we turn to the phenomenon of $D<2$. Corresponding process is known to be persistent or 'with memory'. The generalized Schr\"{o}dinger equation that one may construct for total range $1<D<2$ is degenerated and unphysical as shown in \cite{N2}. So, we consider only the the typical for most markets case of small deviation $D$ from 2. In this case the scale dependence on $\tau$ can be referred to $\mathcal D$. Namely, assuming non-stationary of stochastic process, we rewrite Eq. (\ref{d}) with $D<2$ in the form familiar for standard Brownian walking,
$$\langle d\xi_{\pm}^2(t)\rangle=\pm 2\mathcal D(t,\tau)dt,
$$
defining a generalized scale-dependent diffusion coefficient
\begin{equation}\label{19}
    \mathcal D(t,\tau)=\mathcal D '(t)\tau^{(2/D)-1}.
\end{equation}

Following the lines of Section 3 we find the correspondent time derivative operator is to be given by the same expression as in Eq. (\ref{vstav}) but with $\mathcal D$ now a function given by Eq. (\ref{19}):
 $$
  \frac{\hat d}{dt}=\frac {\partial}{\partial t}+\mathcal V \nabla-i\mathcal D(t,\tau)\Delta.
$$

After this we define the coefficient $\langle\mathcal D\rangle$, that is constant respective to variable $t$ but scale-dependent,
$$\mathcal D =\langle\mathcal D\rangle+\delta\mathcal D (t),
$$
and introduce the complex function $\psi$ from the formula
$$ \psi=Ae^{iS/2m\langle\mathcal D\rangle}.$$

Then $\psi$ is related to the complex velocity as:
$$\mathcal V=-2i\langle\mathcal D\rangle\nabla ln\psi.$$

Further, by steps of Section 4 we come to generalized Schr\"{o}dinger equation:
\begin{equation}\label{vol1}
    \mathcal D\Delta\psi+i\frac{\partial\psi}{\partial t}=\left(\frac{\Phi}{2m\langle\mathcal D\rangle}+\delta\mathcal D(\nabla\ln\psi)^2 \right)\psi.
\end{equation}

Thus the effect of varying diffusion (volatility) amounts to changing the form of potential in the standard Schr\"{o}dinger equation. The non-linear on $\psi$ additive to financial potential responses for 'memory' in prices.
Such a behavior could be of interest in the perspective of a future development of market chaotic dynamics based on the concept of price space fractality and quantum theory.

The second conclusion is to approve that to each time scale (horizon) own Schr\"{o}dinger equation corresponds. On one time horizon variations $\delta D$ are essential (as e.g. intraday volatility oscillations on foreign exchange market \cite{MS}), others not. Under assumption $\delta\mathcal D \ll\mathcal D$, the effect of the additional term and the effect of $\mathcal D$ being a function of $t$ can be treated perturbatively. On gets the equation
\begin{equation}\label{vol2}
    \langle\mathcal D\rangle^2\Delta\psi+i\langle\mathcal D\rangle\frac{\partial\psi}{\partial t}=\left(\frac{\Phi}{2m}-\langle\mathcal D\rangle[\Delta\ln\psi]_0 \delta\mathcal D (t)\right)\psi
\end{equation}
suitable for practical calculations.
\section{Conclusion}
Thus it has been shown that the quantum mechanical description for financial markets is possible on the basis of small number of assumptions, namely fractality (scale invariance) and non-differentiability of price space, and Newtonian character of price dynamics. In the specified framework the price spot is thought by a particle moving under pressure of economic and psychological factors. The economic factors through the cumulative financial potential in the Newton equation, obviously, first of all defines the general conditions for the functioning of market, whereas the potential of '$\psi$-field' driven by Schr\"{o}dinger equation, provides the fractality of trajectory.

We suggest the wave function amplitude is a measurable quantity in the sense that it can be directly estimated from price data. Through it we can get in principle the mathematical description of leading market factors. Studying of wave function evolution perhaps might help in solving problem of price predictability.
\section*{References}

\end{document}